\documentclass[english,aip,journal=jcp,twocolumn,reprint]{revtex4-1}
\usepackage[T1]{fontenc}
\usepackage[latin9]{inputenc}
\usepackage{textcomp}
\usepackage{amstext}
\usepackage{graphicx}
\usepackage{subscript}
\usepackage{lmodern}
\usepackage{color}
\usepackage{enumitem}
\usepackage{tabularx}
\newcolumntype{R}{>{\raggedleft\arraybackslash}X}

\usepackage{amsmath,amssymb}
\usepackage{graphicx}
\usepackage{times,mathptmx} 
\usepackage[pdftex,
  colorlinks=true,%
  urlcolor=blue,%
  linkcolor=blue,%
  citecolor=blue,%
  pdftitle={Phase transitions in water--Xe system},%
  pdfauthor={Vasilii Artyukhov}%
]{hyperref}
\usepackage[all]{hypcap}
\makeatletter

\providecommand{\tabularnewline}{\\}

\@ifundefined{textcolor}{}
{%
 \definecolor{BLACK}{gray}{0}
 \definecolor{WHITE}{gray}{1}
 \definecolor{RED}{rgb}{1,0,0}
 \definecolor{GREEN}{rgb}{0,1,0}
 \definecolor{BLUE}{rgb}{0,0,1}
 \definecolor{CYAN}{cmyk}{1,0,0,0}
 \definecolor{MAGENTA}{cmyk}{0,1,0,0}
 \definecolor{YELLOW}{cmyk}{0,0,1,0}
}

\renewcommand{\fnum@figure}{\textbf{FIG. \thefigure}}
\renewcommand{\fnum@table}{TABLE \thetable}

\usepackage{rotating}
\usepackage[english]{babel}

\makeatother

\begin{document}

\title{Can xenon in water inhibit ice growth? Molecular dynamics of phase
transitions in water--Xe system}

\author{Vasilii I. Artyukhov}

\email{artyukhov@rice.edu}

\selectlanguage{english}%

\affiliation{Department of Materials Science and NanoEngineering, Rice University,
Houston, Texas 77005}

\author{Alexander Yu. Pulver}

\author{Alex Peregudov}

\author{Igor Artyuhov}

\email{igor.artyuhov@bioaging.ru}

\selectlanguage{english}%

\affiliation{Institute of Biology of Aging, Moscow, 125284 Russia}
\begin{abstract}
Motivated by recent experiments showing the promise of noble gases as
cryoprotectants, we perform molecular dynamics modeling of phase transitions
in water with xenon under cooling. We follow the structure and dynamics
of xenon water solution as a function of temperature. Homogeneous
nucleation of clathrate hydrate phase is observed and characterized.
As the temperature is further reduced we observe hints of dissociation
of clathrate due to stronger hydrophobic hydration, pointing towards
a possible instability of clathrate at cryogenic temperatures and
conversion to an amorphous phase comprised of ``xenon + hydration
shell'' Xe$\cdot$(H\textsubscript{2}O)\textsubscript{21.5} clusters.
Simulations of ice--xenon solution interface in equilibrium and during
ice growth reveal the effects of xenon on the ice--liquid interface,
where adsorbed xenon causes roughening of ice surface but does not
preferentially form clathrate. These results provide evidence against the ice-blocker
mechanism of xenon cryoprotection.
\end{abstract}
\maketitle

\section{Introduction}

Ever since the original deciphering of clathrate hydrate structures
by Claussen, Pauling, and others, \cite{1951claussensuggested,*1951claussenerratum,1951stackelbergonthe,1951claussenasecond,1952paulingthestructure}
researchers have been fascinated by the interactions of hydrophobic
gases with water producing these beautiful crystal structures. Natural
gas hydrates are a potential source of energy \cite{2009boswellisgas}
and problem in fossil fuel transport; \cite{2003sloanfundamental}
clathrates could play a role in global climate change \cite{1996kaiholatestpaleocene}
and be used in CO\textsubscript{2} sequestration. \cite{2006parksequestering}
Clathrate hydrates were famously proposed as a molecular explanation
of general anaesthesia. \cite{1961paulingamolecular}

Xenon, a noble gas that readily forms a clathrate hydrate, manifests
distinct physiological actions despite its chemical inertness. Owing
to its strong anesthetic effect, \cite{2004heckerxenona} xenon is
considered to be the best (though expensive) inhalation anesthetic
to date and is officially approved for application as such in Germany
and Russia. Xenon has an antidepressant effect (presumably due to
its anti-NMDA \cite{2009derwallxenonrecent} action) and is used to
treat alcoholism and drug addiction. \cite{2002kornetovapplication}
Xenon shows a protective effect on various organs subject to ischaemia--reperfusion
injury \cite{2005weberthenoble,2005abrainipotentially,2009derwallxenonrecent}
(presumably due to its anti-NMDA effect and interference with many
aspects of intracellular Ca\textsuperscript{2+} homeostasis \cite{2003petzeltprevention}).

An important potential application of xenon is its cryoprotective
action. \cite{1969prehodasuspended,1984rodinstructure,2004shcherbakovmethodfor,2008shelegcardiac}
In order to understand this phenomenon, the physicochemical basis
of interaction of xenon with liquid water, ice and biological molecules
must be understood at the microscopic level. The mechanisms that xenon\textquoteright{}s
antifreeze effect could be ascribed to include: 

\begin{enumerate}[label=(\emph{\roman*}),noitemsep]

\item Trapping of water in clathrate hydrate in competition with
ice crystals formation (akin to traditional cryoprotectants such as
glycerol and DMSO).

\item Vitrification \cite{2004fahycryopreservation} of xenon
solution.

\item Ice-blocker action via kinetic suppression of ice growth
(similar to biological antifreeze proteins \cite{2007pertayagrowthtextendashmelt,2011howardneutron})
by clathrate hydrate forming on ice crystal surface. \cite{1962barrernonstoichiometric}

\item Accumulation of xenon in lipid membranes. \cite{2012yamamotodiffusive,2013bookerbiophysical}

\item Protection against protein thermal denaturation. \cite{1984rodinstructure}

\end{enumerate}

In the present molecular dynamics study we focus on the ``nonbiochemical''
mechanisms (\emph{i}--\emph{iii}) while the interaction of xenon with
biological membranes and proteins is left for follow-up studies. First we investigate 
the structure and properties of homogeneous xenon solution and its behavior under 
cooling below the freezing temperature of water. We observe the formation of 
clathrate phase, with a complex dependence on the concentration and temperature 
of solution, and a signature of another metastable amorphous phase upon deep 
supercooling. Then we study the process of ice growth in the presence of 
dissolved xenon, observing preferential adsorption of xenon on the ice surface but 
no preferential nucleation that could have caused an ice-blocking effect.

\section{Computational details}

\subsection{Simulation method and models}

For xenon, we used Lennard--Jones interaction parameters \cite{1993guillotacomputer}
that were previously successfully used to study hydrophobic hydration
of xenon by Paschek: \cite{2004paschektemperature} $\sigma=3.975\,\AA$, 
$\epsilon/k_{\text{B}}=214.7\,\text{K}$. To simulate water, we used the
TIP5P five-site rigid model. \cite{2000mahoneyafivesite} Cross-species
interaction parameters were derived using the geometrical mean combination
rule. TIP5P was chosen to represent water because it has pronounced
directional hydrogen bonding thanks to its lone-pair virtual sites,
leading to stronger tetrahedral structuring of water, which is important
here because of our interest in the processes of hydrophobic hydration
that is crucially dependent on the local ordering of water molecules
\cite{2004paschektemperature} (especially at low temperatures \cite{2005paschekhowthe}),
and crystal formation. The parametrization of the TIP5P model was
done over a wide range of temperatures and pressures, \cite{2000mahoneyafivesite}
and it correctly reproduces both the anomalous density maximum of
water at 4 \textdegree{}C as well as the melting temperature of ice
Ih ($T_{\text{M}}$ = 275 K as observed in the present work and in earlier
studies \cite{2005vegathemelting,2006vrbkahomogeneous,2005wangmelting,2003nadaanintermolecular}).
Although the thermodynamically most stable ice polymorph is erroneously
predicted by TIP5P to be ice II, \cite{2005vegathemelting} and the
proton-ordered ice Ih structure has lower predicted free energy than
the (correct) disordered ice I, \cite{2003nadaanintermolecular} we
still believe TIP5P to be the most reliable choice of water model
for our purposes.

Molecular dynamics simulations were carried out using the \textsc{gromacs}
package. \cite{2005vanderspoelgromacs} The simulation time step was
2 fs. Coulomb and Lennard-Jones interactions were truncated at a distance
of 0.9 nm (as in the original TIP5P parametrization). Long-range Coulomb
interactions were calculated using the Particle-Mesh Ewald method,
\cite{1993dardenparticle,1921ewalddieberechnung} and a standard long-range
dispersion correction term was applied. \cite{1991allencomputer}
For liquid systems, preequilibration was done using the Berendsen
thermostat, \cite{1984berendsenmolecular} and production runs were
performed using the Nos\'e--Hoover thermostat \cite{1984noseamolecular,1985hoovercanonical}
with a 0.5 ps time constant and the Nos\'e--Parrinello--Rahman barostat
\cite{1981parrinellopolymorphic,1983noseconstant} with a 2.0 ps time
constant. For two-phase systems, production runs used the velocity
rescaling thermostat \cite{2007bussicanonical} and the Berendsen
barostat \cite{1984berendsenmolecular} with a 0.5 ps time constant
for both, to accommodate the total energy and volume change during
phase transition.

\subsection{System preparation and simulation protocols}

\begin{table}

\caption{Duration of simulation runs for homogeneous systems (ns)}
\bgroup
\setlength{\extrarowheight}{3pt}
\begin{tabularx}{\columnwidth}{ l *{6}{R} }
\toprule
$T$ (K) \label{table} & 280 & 270 & 265 & 260 & 255 & 250 \tabularnewline
\hline
1\% & 80 & 80 & 50 & 100 & 100 & >100 \tabularnewline
2\% & 50 & 50 & 50 & 100 & 100 & >100 \tabularnewline
3\% & 30 & 30 & 50 & 100 & 100 & >100 \tabularnewline
4\% & 30 & 30 & 50 & 100 & 100 & >100 \tabularnewline
5\% & 30 & 50 & 100 & 100 & 100 & >100 \tabularnewline
\toprule
\end{tabularx}
\egroup
\end{table}

Liquid systems were prepared by randomly inserting xenon molecules
into a preequilibrated 500-molecule water box to produce a desired
xenon concentration of 1, 2, 3, 4, or 5\% mol.Xe/mol.H\textsubscript{2}O
(meaning 100, 80, 60, 40, and 20 water molecules per xenon, respectively).
The 5\% concentration was chosen as the upper boundary since it still,
in principle, allows each xenon molecule to have its separate hydration
shell (which contains about 20 molecules); at higher concentrations,
there would necessarily be insufficient water to solvate each xenon
molecule, casting doubt on our treatment of the system as a liquid
solution.

The systems were annealed at fixed cell dimensions for a short period
(200 ps) starting at a temperature of 600 K and cooling down to the
target temperature of 250--280 K. After that, they were replicated
to produce $2\times2\times2$ supercells (4000 water molecules, 40--200
xenon) and incubated at the target temperature until equilibrium was
reached. The corresponding run lengths ranged from 30 to 100+ ns (see
Table~\ref{table}). Such large systems and long simulation times were
necessary to collect good statistics for xenon molecules, and because
of the long equilibration times at low temperatures due to very slow
diffusion. At the lowest temperatures, simulation times even this
long sometimes wouldn't produce full convergence of radial distribution
function (RDF) and mean-square displacement (MSD) plots; at the lowest
concentration of 1\%, though the spatial configurations of xenon atoms
were written out as frequently as once every 100 fs (more than sufficient,
given the typical timescale of xenon motion of $\sim$1 ps, as observed
below), RDF plots still contain noticeable statistical noise. 

Two-phase solid-liquid equilibrium simulations were performed with
a $\left( 10 \overline{1} 0\right) $ ice Ih slab, since the prism plane of ice grows 
faster than basal and hence is the most relevant from the viewpoint of
both pure physics \cite{2006cahoongrowthmelt} and biological applications.
\cite{2007pertayagrowthtextendashmelt} The size of the slab
was $5.3\times5.8$ nm, and the thickness was 3.07 nm, corresponding
to four nonprimitive crystal cells, or eight water molecule bilayers
(see figures below); the total number of water molecules in the slab
was 3072. Initial coordinates for ice molecules were taken from the
work of Buch \emph{et al.} \cite{1998buchsimulations} The liquid
layers were taken approximately twice as thick as the slab (6.1 nm)
to accommodate enough xenon molecules and to allow us to study 
longer-range effects of ice on the structure of the solution. Liquid
and solid phases preequilibrated at 275 K were brought in contact
and relaxed using steepest descent energy minimization with a maximum
force criterion of 100 (kJ/mol)/nm, which was found to be sufficient
to remove any bad molecular contacts at the interface while still
ensuring that atomic coordinates still correspond well to the 270--275
K temperature of subsequent production runs, thus providing smooth
continuation of molecular dynamics. 

The results reported herein were mostly produced at a simulated pressure
of 1 atm irrespective of xenon concentration in the systems. While
xenon clathrate hydrate is stable under these conditions or close
(its experimental dissociation pressure value is 2.5 bar at 5 \textdegree{}C,
1.5 bar at 0 \textdegree{}C, and 1 bar at -3.6 \textdegree{}C\cite{1974ewingdissociation}),
 these concentrations are 
considerably above equilibrium at a pressure of 1 atm, which may create an 
excessive driving force for clathrate formation. Extrapolation of empirical xenon solubility data\cite{Note1} 
to our temperature range suggests that our concentrations should correspond 
to ``equilibrium'' (barring formation of solid phases) pressures on the order 
of tens to few hundred atmospheres, which is, in fact, a much milder driving force 
than used in most if not all recent MD studies on clathrate formation. We never observed
separation of xenon gas phase in our simulations; in contrast, such
separation was observed to occur spontaneously in molecular dynamics
simulations of methane hydrate at 100 MPa; \cite{2006nadagrowthmechanism}
note that the dissociation pressure of methane hydrate is only $\sim$5
MPa, and the model system was smaller than ours (1472 water molecules),
although the gas concentration was higher (256 molecules, or 17\%
mol.Me/mol.H\textsubscript{2}O). Further, our tests on two-phase
systems at 10 atm did not produce any noticeable change in the
equilibrium characteristics that would suggest the need for computationally
expensive pressure dependence studies, and the decomposition temperature
of xenon clathrate hydrate did not depend on the concentration (see
discussion below). In summary, this leads us to believe that this
choice of simulation conditions does not significantly affect the
quality of our results (and, in fact, allows us to draw additional
important conclusions). Physically, this corresponds to the approximation
that the pressure dependence of the balance between gas-phase and
dissolved xenon molecules, which determines the concentration of xenon
in solution, is primarily governed by the change in the chemical potential
in the gas phase, while the structure and thermodynamics of the dissolved
phase are, to a reasonable degree, unaffected by pressure on the relevant
magnitude scale.

\section{Results}

\subsection{Liquid-phase simulations}

To study the effects of xenon on the freezing of water, we performed
simulations of five model systems with different xenon concentration
at the temperatures of 280, 270, 265, 260, 255, and 250 K (production
run durations are listed in Table~\ref{table}) . The initial motivation had
been to study the dependence of xenon-xenon interactions, as manifested
in radial distribution function (RDF) plots, and xenon diffusion rate
determined from the slope of plots of mean square displacement (MSD)
over time, on the temperature and concentration of xenon in water
solution below the freezing point.

\begin{figure}[b]
\includegraphics[width=1\columnwidth]{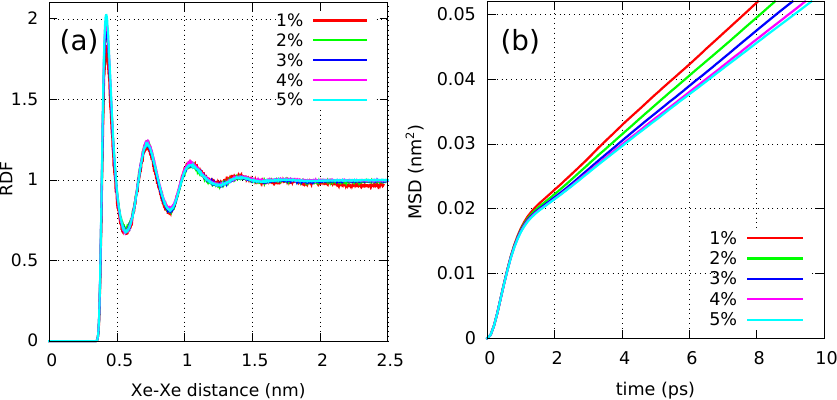}

\caption{\textbf{Xenon solution above phase transition temperature.} Plots
of (\textbf{a}) Xe--Xe radial distribution function (RDF) and (\textbf{b})
Xe mean-squared displacement (MSD) at 280 K. \label{fig:280K}}
\end{figure}

\subsubsection{Structure and dynamics of xenon solution}

\textbf{Figure~\ref{fig:280K}} shows the plots of Xe--Xe RDF and
Xe MSD time dependence for the concentration range studied in the
present work at a temperature of 280 K. It can be seen that the structure
of all five systems is identical to a good precision (the corresponding
Xe--O RDF are also indistinguishable and hence not shown). By 
inspecting the structures, we observe that the first
three peaks in the RDF correspond to the following three cases of relative position
of xenon molecules:
\begin{enumerate}
\item \emph{contact solute pair} (CSP): direct contact of xenon molecules
(hydrophobic association in a common hydration shell; this peak was
previously shown to be enhanced with increasing temperature \cite{2004paschektemperature})---the
maximum of the peak roughly corresponds to twice the radius of xenon
atom ($\sim$0.4 nm); 
\item \emph{solvent-separated solute pair} (SSSP): face-sharing of water
cages containing two xenon molecules (water molecules ``sandwiched''
between xenon molecules)---the maximum corresponds to the clathrate
hydrate structure (see below);
\item contact of water cages: hydration shells of the two xenon molecules
have no common molecules; a hydrogen bond connects a vertex of one
with a vertex of the other---like, for example, the two 5\textsuperscript{12}
cages in the clathrate unit cell. Geometrically, this corresponds to $d_{\text{\text{Xe\textendash Xe}}}$
{[}1.1 nm from Xe-Xe RDF{]} $\approx2d_{\text{Xe\textendash O}}$
{[}0.76 nm from Xe--O RDF{]} + $d_{\text{O\textendash O}}$ {[}0.27
nm from O--O RDF in pristine water{]}. We propose the term \emph{contact
cluster pair} (CCP) to describe this configuration.
\end{enumerate}
However, the dynamics of the systems shows a more complex two-timescale
behavior. The first part of the MSD time dependence is completely
identical for all systems, while the second part demonstrates a slow-down
of xenon diffusion with increasing concentration. This indicates that
there two modes of xenon atom motion in the solution:
\begin{enumerate}
\item fast confined oscillatory motion of xenon inside the hydration cage
with a characteristic time of about 1 ps and a 
squared amplitude of about (0.112 nm)\textsuperscript{2},
showing no visible dependence on the structure of the medium surrounding
the hydration shell, and
\item slow drift of the cluster (xenon molecule plus its hydration
shell) as a whole in the surrounding solution with occasional hopping 
of xenon from shell to shell (see Movie S1\cite{Note2}),
on the timescale of over 1 ps.
\end{enumerate}
The dependence of the second type of motion on the concentration can
be attributed to their collisions decreasing the effective ``mean free
path'', or by formation of more massive cluster
aggregates (potential clathrate hydrate nuclei). 

\begin{figure}[t]
\includegraphics[width=1\columnwidth]{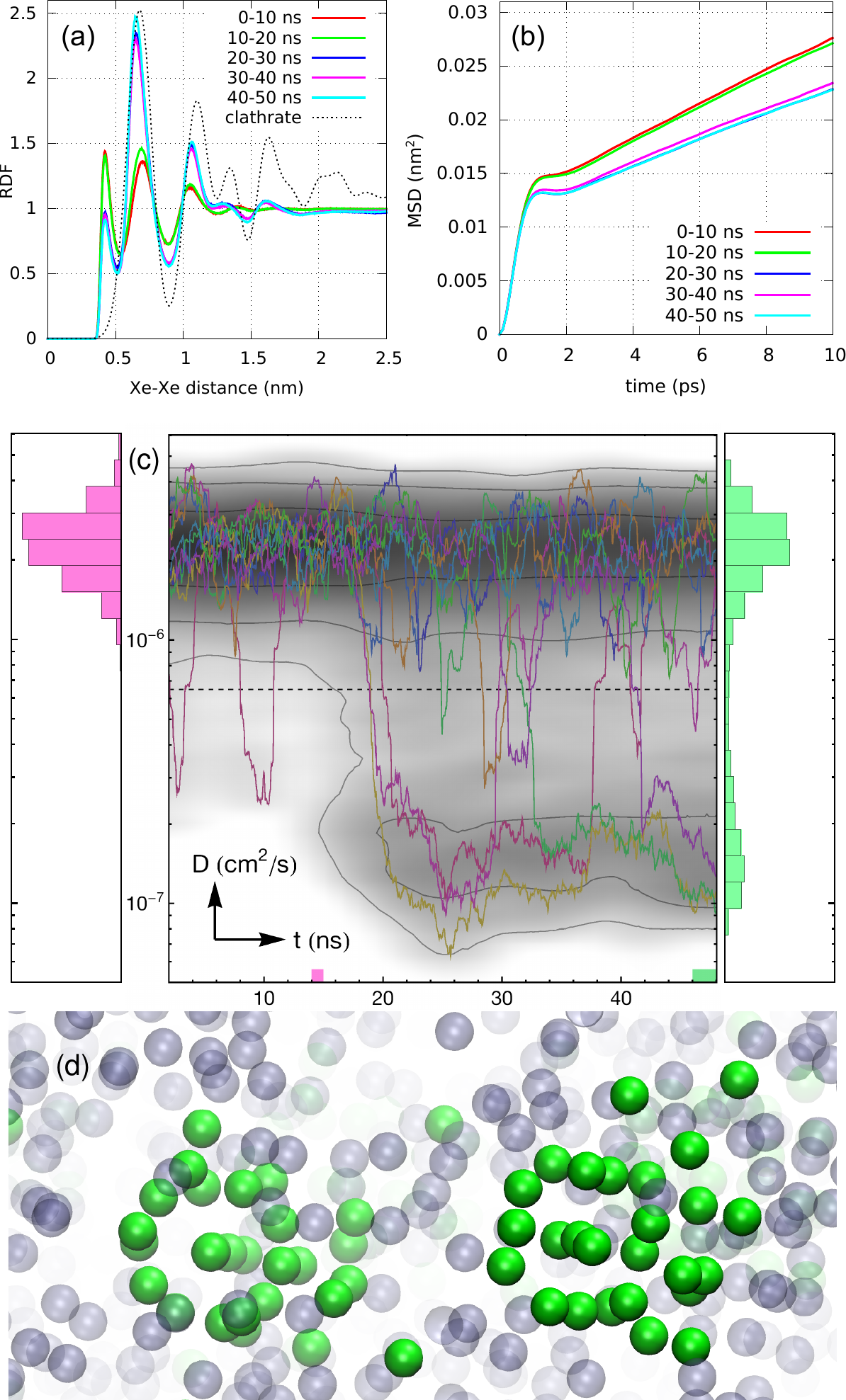}

\caption{\textbf{Phase transition upon cooling.} Sequential snapshots of (\textbf{a})
Xe-Xe RDF and (\textbf{b}) Xe MSD for the 5\% mol.Xe/mol.H\textsubscript{2}O
system at 270 K over 10-ns intervals. A phase transition is observed after the first 20 ns.
(\textbf{c}) Atom-by-atom analysis of mobility. Colored lines 
show the trajectories of diffusion coefficient $D\left(t\right)$ for a representative 
sample of 10 atoms. The background shows the full $D$ distribution for all atoms. 
The histograms on the left and right correspond to $t=14\div15$ ns and $t>46$ ns, 
respectively. (\textbf{d}) A close-up image of two crystallites around $t\sim47$ ns.
The green atoms have an instantaneous $D$ below the dashed horizontal line in (\textbf{c}).
See \textbf{Movie S2}\cite{Note2} for an animation of the entire run. \label{fig:270K}}

\end{figure}

\subsubsection{Cooling of xenon solution}

At temperatures of 270 K and below, in systems with highest xenon
concentrations, a phase transition was observed, leading to a structure
radically different from liquid solution. \textbf{Figure~\ref{fig:270K}}
shows the RDF (\textbf{a}) and MSD (\textbf{b}) plots for the 5\% mol.Xe/mol.H\textsubscript{2}O
system at 270 K over five sequential 10-ns simulation intervals. It
can be seen that during the initial 20 ns, the system was in a liquid
state same as above: the RDF peaks have the same positions; the suppression
of the leftmost CSP peak compared to 280 K is a manifestation of the
hydrophobic dissociation with decreasing temperature; it results from
the negative entropy of hydration, \cite{2004paschektemperature}
which makes it favorable for water molecules at low temperatures to
become a part of a hydration shell, forcing the neighboring xenon
molecules apart to increase the ``surface area'' of hydration. The
dynamics during the first 20 ns also corresponds to that seen in \textbf{Figure~\ref{fig:280K}},
with fast motion on a 1 ps time scale with a (0.122 nm)\textsuperscript{2}
squared amplitude and a slower mode over longer time scales, but with
a slope smaller than at 280 K, obviously, due to the lower temperature.

However, after the initial 20 ns, the system underwent an abrupt phase
transition, manifesting itself both in the structure and 
dynamics of the system. In the RDF plot (\textbf{Figure~\ref{fig:280K}(a)}), it can be seen that
the height of the first peak dropped sharply (below 1, indicating
effective repulsion of xenon molecules), while the second peak became
much more intense and shifted slightly to the left; the third peak
moved to the right, and some previously unseen ordering appeared at
larger distances.

To analyze the changes in the system we calculated the diffusion coefficient $D$ 
every 20 ps for each atom using 200-ps MSD fits. We then computed the running average 
over 100 consecutive values to smoothen the fluctuations. The results are plotted in \textbf{Figure~\ref{fig:270K}(c)}. 
The density and contour maps in the background represent the distribution of atomic 
mobilities over time. The colored solid lines show the $D\left(t\right)$ trajectories 
of a representative random sample of 10 atoms. It can be seen that the system starts in 
a relatively homogeneous state with most atoms having $D\sim 2\times10^{-6}$ cm\textsuperscript{2}/s 
but after about 15 ns a number of them make a transition to a state with an order-of-magnitude 
lower mobility, $D\sim10^{-7}$ cm\textsuperscript{2}/s. The left and right histograms in \textbf{Figure~\ref{fig:270K}(a)} 
further illustrate the qualitative changes in atomic mobility distributions before this event 
and at the end of our MD trajectory, with the striking bimodal distribution seen in the right chart.

To better understand the nature of the phase transition 
we visualized the trajectory coloring the atoms according to their instantaneous diffusion 
coefficient. In \textbf{Movie S2}\cite{Note2}, atoms having a diffusion coefficient lower than an arbitrary 
cutoff shown by the dashed line in \textbf{Figure~\ref{fig:270K}(c)} are highlighted in green. 
In the first $\sim10$ ns of the trajectory, two events of short-lived cluster formation are observed. 
These are seen in \textbf{Figure~\ref{fig:270K}(c)} as two small ``chunks'' below the dashed line 
on the left. This suggests that our conditions are very close to reversibility. Then, after 15 ns, 
a supercritical cluster forms and continues to grow, followed shortly by a second one, until the end of the run. 
The final structure of the two crystallites, shown close-up in \textbf{Figure~\ref{fig:270K}(d)}, visibly corresponds to 
the cubic sI clathrate lattice of xenon 
hydrate (\textbf{Movie S2}\cite{Note2}), and agrees well with the experimental lattice
constant value of 1.182 nm. Even though a strict assignment 
of crystal structure is ill-posed for such a small cluster, and clathrate nucleation 
may involve ``amorphous'' clathrate-like precursors,\cite{2010jacobsonamorphous} the RDF for ideal sI  
lattice (\textbf{Figure~\ref{fig:270K}(a)}, dashed line, with Gaussian broadening applied 
to match the peak height) clearly shows that the trends are consistent with 
clathrate formation. In this context, the ``migration''
of xenon molecules from the first RDF peak (CSP) to the second (SSSP)
is a result of all guest molecules in the clathrate hydrate sharing
cage walls with their neighbors (no immediate xenon--xenon contacts).
The shift of the third (CCP) peak to the right may be due to the fact
that the T-cage in sI clathrate hydrate (tetrakaidecahedron---6\textsuperscript{2}5\textsuperscript{12})
is actually somewhat larger than the dodecahedral cage, which is the
preferred shape of hydration shell in the liquid state (as indicated
by, e.g., NMR studies of xenon clathrate hydrate growth; \cite{1995pietrassmonitoring}
integration of the first peak of Xe--O RDF {[}not shown{]} in the
liquid systems also produces an average number of molecules in the
hydration shell of 21.5, consistent with the predominance of dodecahedral
cages and in contrast to the 23.0 value of ideal clathrate hydrate
that has 8 xenon molecules per 2 5\textsuperscript{12} and 6 6\textsuperscript{2}5\textsuperscript{12}
cages in the unit cell). 

It should be noted that, although only a minor fraction of xenon molecules
get converted to the clathrate phase, the stability of crystalline
structure with respect to thermal motion of atoms results in the fact
that this structure is very markedly seen in the RDF. At the same time, 
no detectable changes are observed in the water O--O nor Xe--O RDF. This is understandable 
as only a small fraction of water converts into clathrate, and even in solution, Xe is 
always surrounded by a polyhedral cage of water molecules. As for the diffusion
rate of xenon, the quantitative differences are much less pronounced
because of the low fraction of encaged xenon molecules, but the transition
is also similarly sharp, and the parameters of the system reached
the new equilibrium almost immediately (on the order of 1 ns). 

\begin{figure}[t]
\includegraphics[width=1\columnwidth]{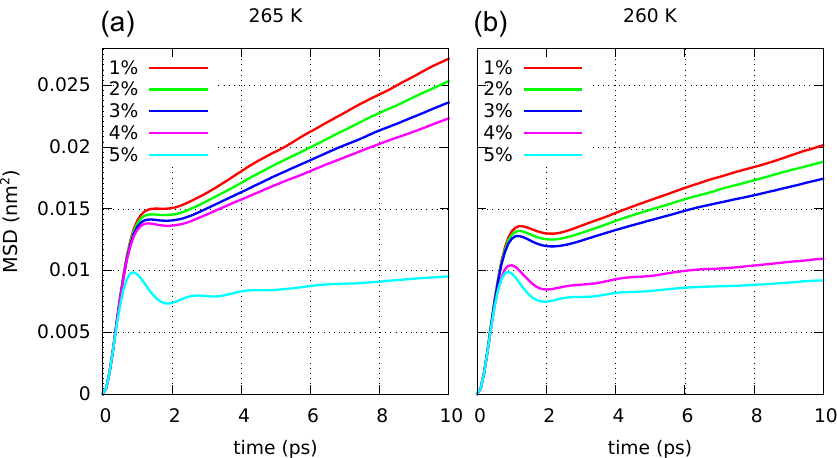}

\caption{\textbf{Concentration dependence of clathrate formation temperature.}
MSD plots for all systems at (\textbf{a}) 265 K and (\textbf{b}) 260
K showing a sharp decrease in mobility associated with phase transition 
above 4\% and 3\% concentration at the respective temperatures. \label{fig:MSD-c-T}}

\end{figure}

\begin{figure*}[t]
\includegraphics[width=17cm]{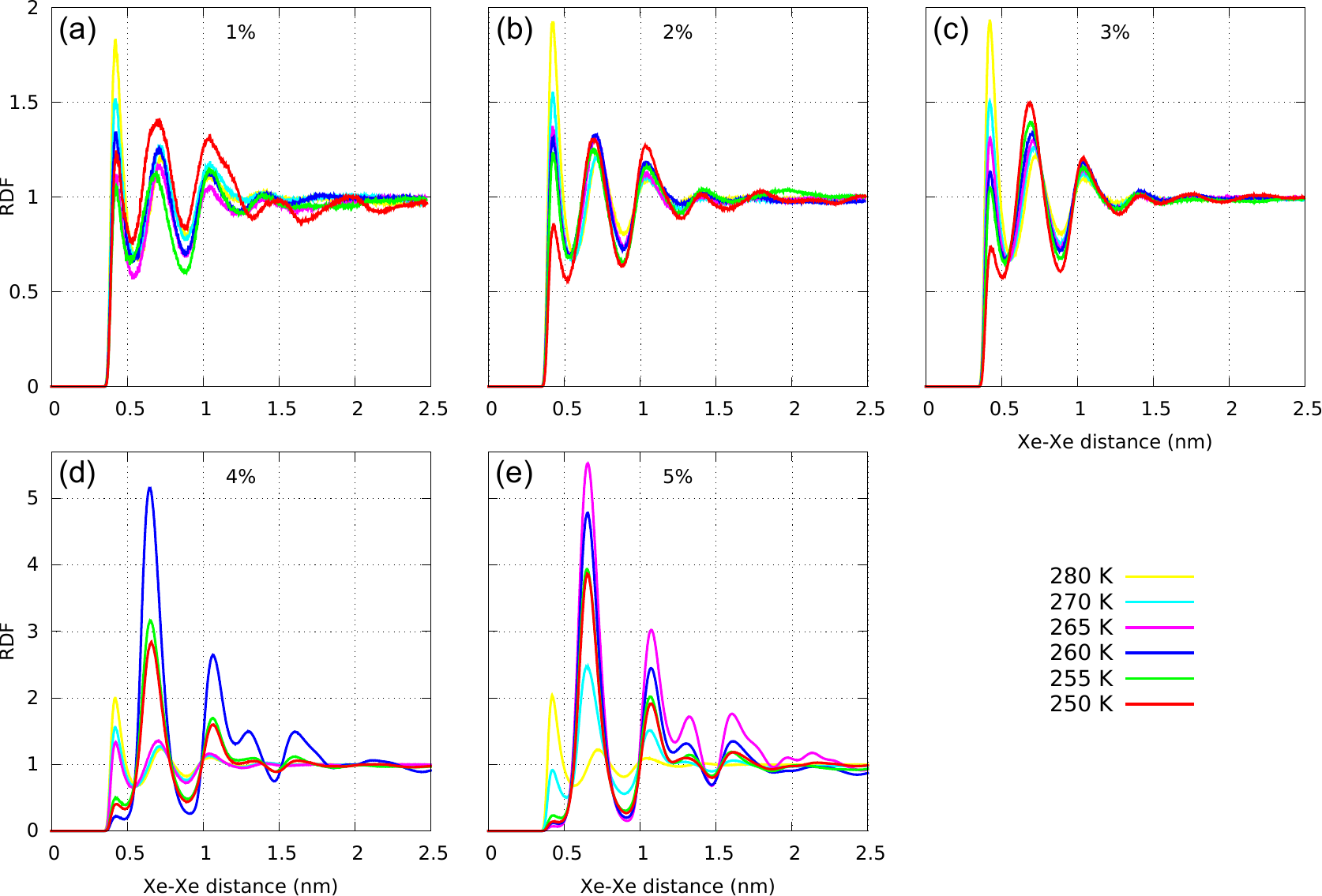}

\caption{\textbf{Formation of clathrate hydrate.} Temperature dependences of
RDF for Xe concentrations of (\textbf{a--e}) 1--5\% mol.Xe/mol.H\textsubscript{2}O.
\label{fig:RDF-c-T}}

\end{figure*}

The change in the dynamical characteristics of the systems upon clathrate
hydrate formation is more clearly seen in \textbf{Figure~\ref{fig:MSD-c-T}},
where MSD plots are shown for different system compositions at the
temperatures of (\textbf{a}) 265 K and (\textbf{b}) 260 K, for which the transition from the
liquid-like to clathrate state is the most pronounced (as demonstrated
by the relative height of the second RDF peaks in \textbf{Figure~\ref{fig:RDF-c-T}}). It can
be seen that at 265 K, only in the 5\% mol.Xe/mol.H\textsubscript{2}O
system clathrate hydrate formation did occur, while at 260 K, it formed
in both the 5\% and the 4\% systems. The following observations can
be made regarding the dynamics of xenon in the systems:

In the ``fast'' part of the MSD curves, there are virtually no differences,
meaning that a xenon atom inside its hydration shell is completely
unaffected by the environment outside the water cage. Upon formation
of the crystalline phase, the amplitude of this oscillation decreases
from (0.122 nm)\textsuperscript{2} at 280 K (\textbf{Figure~\ref{fig:270K}})
to (0.114 nm)\textsuperscript{2} at 270 K and further down to (0.105--0.100
nm)\textsuperscript{2} at even lower temperatures, where the number
of xenon molecules in the clathrate phase is larger, and hence, their
statistical contribution is greater. This reduction is probably linked
to the greater rigidity of the clathrate lattice compared to liquid
water, which prevents the hydration shell from deformations that could
accommodate larger xenon displacements (xenon molecules ``bounce off''
cage walls without deforming them).

In the ``slow'' part, the slope of MSD curves is substantially decreased
compared to the liquid phase. In the first approximation, this decrease
can be considered proportional to the fraction of xenon in the clathrate
phase, since the motion of the latter is that of the crystallite as
a whole---orders of magnitude slower (at least, in the absence of
vitrification of solution). For systems where no clathrate formed,
extrapolation of the ``slow'' parts to $t=0$ results in almost similar
values of (0.105--0.110 nm)\textsuperscript{2}, which is close to
the above-reported amplitude of xenon molecule motion in the cages
of clathrate hydrate.

At low temperatures, traces of fast motion can be seen against the
background of the slow drift. These indicate the former
indeed has the character of oscillation. The period of these oscillations
in the crystalline phase is about 2.0 ps; in the liquid phase, it
is somewhat greater---about 2.2 ps---which is probably again linked
to the greater flexibility of the hydrogen bond network in the liquid
compared to the clathrate.

As seen in \textbf{Figure~\ref{fig:RDF-c-T}}, crystalline clathrate
hydrate formation is only observed in the most concentrated 5\% (270
K and below) and 4\% mol.Xe/mol.H\textsubscript{2}O (260 K and below)
systems, while at lower concentrations, no hints of crystalline phase
are seen in the RDF (except maybe for a small increase of second peak
intensity for the 3\% system at 255 and 250 K; visual inspection revealed
some signs of emerging ordering but no well-identifiable crystallites).
However, it is not entirely clear if this absence of clathrate hydrate
is due to the concentration being insufficient for homogeneous nucleation,
or simply because of kinetic restrictions not allowing for a detectable
amount of crystalline phase to grow over the limited duration of the
simulation.

As for the 1--3\% systems that remain liquid over the entire temperature
range, a steady monotonous behavior is observed in RDF as a function
of the decreasing temperature. The intensity of the first CSP peak
gradually decreases, while the SSSP and CCP peaks grow. This is a
manifestation of the hydrophobic dissociation, as described above.
In addition, the RDF minima become deeper, and higher-order maxima
(4th, 5th, and 6th order, up to $\sim$2.2 nm) become clearly distinguishable.
This pronounced structuring of the systems without any detectable
crystallinity, as well as the associated collective dynamics in these
and similar systems, deserve separate attention outside the boundaries
of this work.

It is also interesting to note that in the 4\% and 5\% systems, there
is actually less clathrate hydrate formed in the end of simulations
at temperatures below the onset of homogeneous nucleation (255 and
250 K for 4\%, 260 K and below for 5\%, see \textbf{Figure~\ref{fig:RDF-c-T}}).
While this could possibly be explained by slower diffusion rate not
allowing enough xenon molecules to drift towards the growing nuclei
of clathrate phase to reach its equilibrium mass during the limited
time of simulation, this explanation appears unlikely since we did
assure full convergence in the RDF for these systems. We therefore
find it more plausible that this effect is a consequence of low-temperature
hydrophobic dissociation: there exists an optimal temperature at which
xenon clathrate hydrate is most stable (e.g. 265 K for the 5\% system,
\textbf{Figure~\ref{fig:RDF-c-T}(e)}), above which it becomes less
stable due to thermal decomposition, and below which it is ``pulled
apart'' as a result of entropy-driven effective repulsion of xenon
molecules (see the RDF plots for the 1--3\% systems, \textbf{Figure~\ref{fig:RDF-c-T}(a--c)}). 
Although the entropy of the molecules that compose
a crystal is much lower than in the liquid state, there is only 5.75
water molecules per each xenon in the clathrate versus an average
of 21.5 in the hydration shell of xenon in the liquid phase, meaning
that despite that the water molecules in the latter case are thermodynamically
of ``poorer quality'' (higher entropy---though, importantly, still
lower than in bulk water), CCP configuration could still be more favorable
because of the sheer quantity of low-entropy water molecules. To put
it in other terms, the hydrophobic solubility of xenon in low-temperature
water increases with cooling faster than the thermal stability of
clathrate hydrate.

The fact that in the 5\% mol.Xe/mol.H\textsubscript{2}O system at
270 K, the first two nucleation events were unsuccessful and it took 
about 23 ns for the first stable crystal nuclei to form, as well as
the low amount of xenon that converted to the clathrate phase, suggests
that these system parameters (composition and temperature) are exceptionally
close to the balance between clathrate growth and dissociation. In
all other cases, clathrate hydrate formation proceeded right from
the start of the simulation---even at the preequilibration stage,---and
the growth of the crystal occurred very rapidly, despite the fact
that even the highest xenon concentrations used in the present work
are still several times smaller than that in pure clathrate hydrate
($\le$ 5\% versus 17.4\% mol.Xe/mol.H\textsubscript{2}O, respectively).
It is also important that clathrate nucleation proceeds much faster
than the homogeneous nucleation of ice (which was never observed in
our simulations, and is generally a notoriously difficult process
to capture with molecular dynamics simulations \cite{2002matsumotomolecular}).
This is because xenon molecules with their well-defined hydration
shells enforce a preferential orientation for the surrounding water
molecules via sandwiching \cite{2009walshmicrosecond} and ``inverse
solvation'' \cite{2010matsumotofourbody}, which radically decreases
the variety of available equiprobable amorphous local structures that
allows pristine water to stay in a supercooled state for macroscopic
periods of time.

\begin{figure}[t]
\includegraphics[width=1\columnwidth]{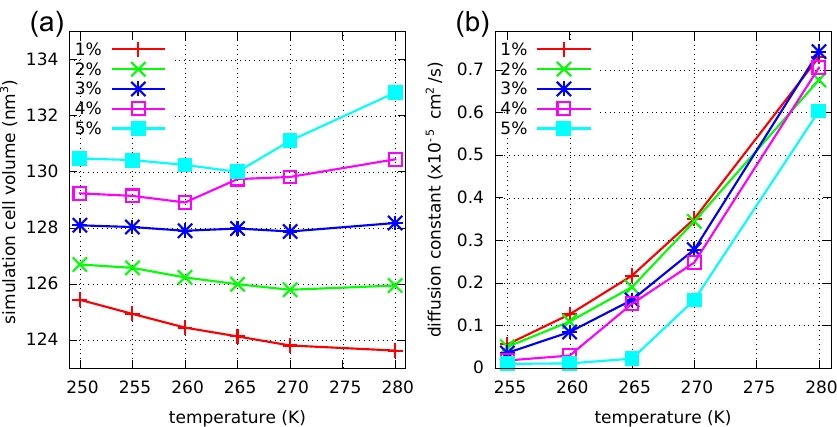}

\caption{\textbf{Thermodynamic signatures of phase transition.} Temperature
dependences of (\textbf{a}) simulation cell volume and (\textbf{b})
xenon diffusion coefficient as a function of system composition. \label{fig:DandVofT}}

\end{figure}

Finally, \textbf{Figure~\ref{fig:DandVofT}} shows the plots of xenon
diffusion coefficients (\textbf{a}) and specific volume (\textbf{b}) of the systems as a function
of the temperature (not shown for $T$ = 250 K because the duration
of our simulations did not allow us to reach full convergence of this
parameter). In both plots, clathrate formation manifests itself in
the form of discontinuities at 5\%/265(270) K and 4\%/260 K; in particular,
the conversion of xenon to the clathrate phase is associated with
an appreciable volume decrease (a point to be recalled below). Another
interesting observation is that, even though water shows anomalous
thermal behavior of density in the studied range of temperatures (density
decreases with temperature), the presence of xenon appears to oppose
this tendency, and for the 3\% system, the volume is nearly constant
throughout the entire 250--280 K range, while at higher concentrations,
when a large fraction of water molecules are part of hydration shells
(above the onset of clathrate formation), this effect even dominates
over the liquid water behavior. While the practical consequences of
this effect are unclear due to the unusual system composition, it
is a further interesting manifestation of the negative entropy change
associated with hydrophobic hydration. \cite{2004paschektemperature,2004paschekheatcapacity}

\subsection{Two-phase simulations}

\begin{figure*}
\includegraphics[width=17cm]{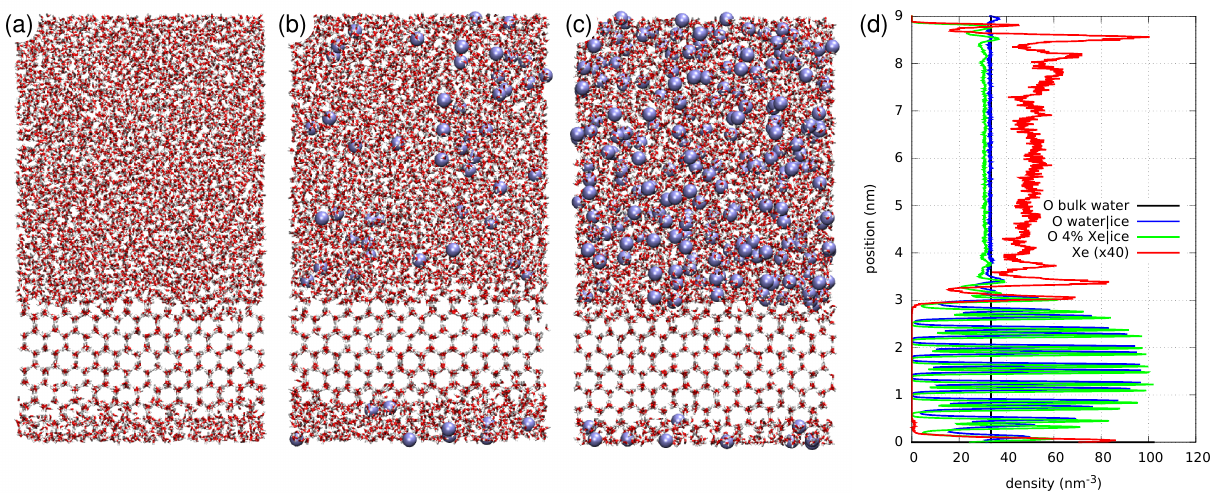}

\caption{\textbf{Phase equilibrium simulations.} Snapshots of two-phase systems,
at the corresponding equilibrium $T_{\text{eq}}$ temperatures, containing
ice and (\textbf{a}) pure water at 275 K, (\textbf{b}) 1\% mol.Xe/mol.H\textsubscript{2}O
solution at 274 K, and (\textbf{c}) 4\% mol.Xe/mol.H\textsubscript{2}O
solution at 271 K. (\textbf{d}) Structure of ice--liquid interface: Xe and O atom density
profiles for systems (\textbf{a}) and (\textbf{c}). \label{fig:2phase-eq}}

\end{figure*}

Two-phase simulations of ice crystals in xenon-rich water solution
were undertaken in order to investigate how xenon affects the growth
of ice, and to assess how our above findings relate to heterogeneous
crystal growth. The simulations were carried out for three systems
with ice slabs embedded in (\emph{i}) neat water, (\emph{ii}) 1\%
mol.Xe/mol.H\textsubscript{2}O liquid solution (low xenon concentration)
and (\emph{iii}) 4\% mol.Xe/mol.H\textsubscript{2}O (high xenon concentration).
The latter concentration was chosen so as to ensure that no clathrate
hydrate forms at our target temperatures (above the homogeneous nucleation
point of 260 K---\textbf{Figures \ref{fig:RDF-c-T}(d) and \ref{fig:DandVofT}}). 
Snapshots of the systems are shown in \textbf{Figure~\ref{fig:2phase-eq}}.

\subsubsection{Equilibrium structure of the interface}

Equilibrium coexistence temperature $T_{\text{eq}}$ values were obtained
via temperature bracketing. Growth/melting of ice was monitored by
the change in the potential energy. The resulting values of $T_{\text{eq}}$
were: 275 K for pristine water; 274 K for the 1\% Xe system; 271 K
for the 4\% Xe system. It can be seen that the deviation of $T_{\text{eq}}$
from 275 K is directly proportional to the amount of xenon in the
liquid phase, i.e., conforming to the normal cryoscopic law. Curiously, 
this suggests that our homogeneous 5\%/270 K run (\textbf{Figure~\ref{fig:270K}}) 
was close to equilibrium with \emph{both} clathrate and ice. Increasing
the pressure to 10 atm did not shift the equilibrium temperatures,
confirming the physical soundness of our choice of $P=1$ atm for
the bulk of our calculations reported above. 

Visual inspection showed that the presence of xenon did affect the
structure of ice surface. In pure water runs, the surface of ice slab
was generally smooth down to the length of one unit cell in the corresponding
dimension. However, the heavy xenon molecules were frequently seen
to plunge into the surface layers of the slab, sometimes up to two
crystal cells deep. This created a roughened surface with disordered
liquid-like local structure of water molecules. The residence times
of xenon molecules were up to about several ns, after which they would
be ejected back into the surrounding solution, and the protrusions
would be healed, restoring the perfect crystalline structure of ice. 

\textbf{Figure~\ref{fig:2phase-eq}(d)} shows Xe and O atom density profiles
for the ice/4\% mol.Xe/mol.H\textsubscript{2}O system at its equilibrium
temperature of 271 K (the xenon plot has been multiplied by 40 to 
compensate for the difference in total atom number to facilitate visual 
comparison). The abscissa corresponds to the distance along the \emph{z}
direction perpendicular to the ice slab surface. The statistics were
accumulated over 10 ns (1000 configurations). 

It can be seen that water molecules are in a crystalline state at
0--3 nm (forming pairs of crystal planes), are uniformly distributed
in the liquid region of the system (4--8 nm), and form quasi-liquid
(or quasi-crystalline) interface layers at 3--4 nm and 8--9 nm. The
distribution of xenon is also more or less uniform---to the extent
of confidence allowed by the poorer statistical sampling---in the
liquid layer (absence of clathrate hydrate was confirmed visually
and by RDF plots), and exhibits 3 or 4 pronounced peaks in the interface
layer with intensities of up to $\sim$1.5--2 times that in the liquid
phase. The Xe peak closest to the surface is located between two O
crystal plane peaks, and the liquid-phase peaks of Xe and O coincide.
Inspection of trajectories 
reveals that the ice/solution interface contains some adsorbed xenon
molecules that stay there for a duration of about 3--5 ns and then
drift away back into the liquid part of the system; at any instant,
only on the order of 10 molecules are adsorbed on the $\sim$31 nm\textsuperscript{2}
surface of the simulation slab cell---a concentration that is insufficient
to cause clathrate hydrate nucleation, even though it is enough to
depress the melting temperature by 4\textdegree{}. 

The origin of Xe concentration peaks can be understood basing on the
difference between ice and liquid water densities. Since the density
of interfacial water is lower than the bulk value, it can more easily
accommodate guest xenon molecules, similarly to what has been previously
observed in simulations of hydrophobic hydration in deeply supercooled
water. \cite{2005paschekhowthe} On the other hand, formation of xenon
clathrate hydrate is associated with the opposite sign of density
change (densification), which suggests that nucleation of xenon clathrate
hydrate should not be more favorable on the ice/water interface than
in the bulk \cite{2009lisurfaceinduced} (see also the next section). 

Finally, comparison of O atom density profiles in \textbf{Figure~\ref{fig:2phase-eq}(d)}
for the cases of pristine water and water with xenon shows an unperturbed
structure, except for a slight enhancement of the third peak in the
quasi-liquid layer, $z=4$ nm (the differences in the topmost part
of the plot are artifacts coming from the different periodic boundary
conditions). This peak should be attributed to water molecules that
solvate xenon adsorbed at the interface. 

The present findings for the prism face of ice can, with reasonable caution,
be expected to generalize to other ice surfaces, e.g., basal. Because of the 
lower relative density of ice, the quasi-liquid layer at its surface is expanded, 
and provides a favorable environment for relatively large xenon molecules, 
thus causing them to preferentially adsorb in this layer. However, additional 
simulations will be needed to fully confirm this.

\begin{figure*}
\includegraphics[width=17cm]{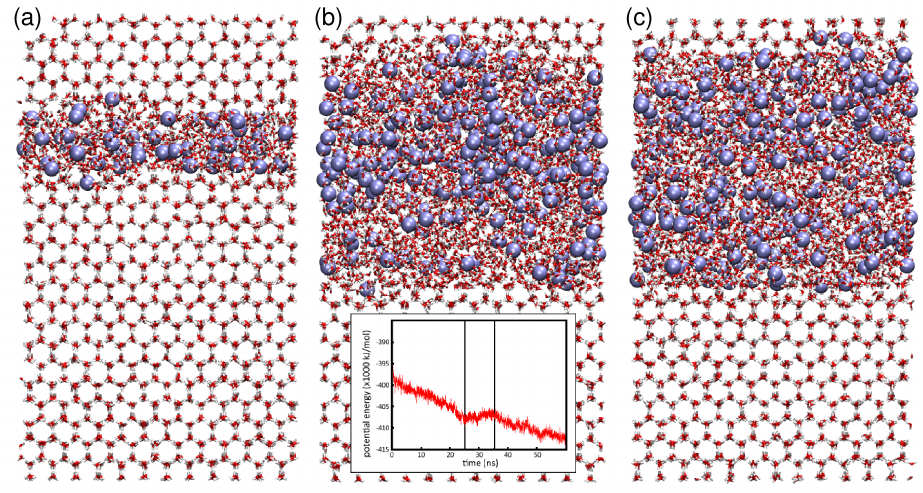}

\caption{\textbf{Freezing of xenon solution.} Final structures of two-phase
systems: (\textbf{a}) 1\% mol.Xe/mol.H\textsubscript{2}O, 270 K,
1 atm; (\textbf{b}) 4\% mol.Xe/mol.H\textsubscript{2}O, 268 K, 1
atm; (\textbf{c}) 4\% mol.Xe/mol.H\textsubscript{2}O, 268 K, 10 atm.
The inset in (\textbf{b}) shows the potential energy profile with
a plateau that corresponds to clathrate hydrate formation in solution (see also Movie S3\cite{Note2}).
\label{fig:freezing}}

\end{figure*}

\subsubsection{Ice growth in the presence of xenon}

After equilibration at the corresponding $T_{\text{eq}}$ values, the temperature
of our systems was decreased to 270 K for 1\% and to 268 K for 4\%
mol.Xe/mol.H\textsubscript{2}O. Although the computational expense
of the simulations prevented us from doing an extensive systematic
investigation, the following behavior was observed:

\emph{1\%, 270 K.} The ice slab smoothly grows, expelling xenon
into the remaining liquid region. The growth completes in a state
when most of the water is in the form of ice, with a thin amorphous
(no crystalline clathrate hydrate) layer containing all the xenon
(\textbf{Figure~\ref{fig:freezing}(a)}, $t=120$ ns).

\emph{4\%, 270 K, 1 atm.} After $\sim$25 ns of ice front propagation,
the concentration of xenon in the liquid layer built up sufficiently
to result in the nucleation of a clathrate hydrate seed ($\sim$10
ns), which then proceeded to grow forming a crystal parallel to the
ice slab surface. The resulting structure ($t=60$ ns) is shown in
\textbf{Figure~\ref{fig:freezing}(b)}, and the corresponding potential
energy plot is shown in the inset. The clathrate hydrate appears as
a relative increase of xenon molecule density slightly above the middle
of the liquid layer; it is more clearly visible in the molecular dynamics
trajectory available in \textbf{Movie S3}.\cite{Note2} It must be stressed that nucleation
started in the middle of the liquid layer and not at the surface of
the ice slab, consistent with our reasoning above based on the density
of the interfacial quasi-liquid layer and of clathrate hydrate. Ice
surface propagation stopped at $\sim$45 ns, although the potential
energy continued to decrease, most likely, because of clathrate growth
or rearrangement.

\emph{4\%, 270 K, 10 atm}. The growth rate of ice was similar to
the 1 atm simulation, and the growth stopped at $\sim$45 ns with
the same ice slab thickness. However, no detectable clathrate hydrate
fragments were formed even after $t=60$ ns (\textbf{Figure~\ref{fig:freezing}(c)};
the distribution of xenon in the liquid layer is uniform). This behavior
appears to be in disagreement with what should be expected basing
on the negative volume change upon clathrate hydrate formation, but
may be simply due to the stochastic nature of crystal nucleation.
Statistics over multiple freezing trajectories would be needed to
clarify this issue.

In summary, we have observed that during ice growth in xenon water
solution, xenon alters the structure of ice surface making it rougher
than in pure water, and that this effect reduces the melting temperature
of ice in agreement with the cryoscopic law. In all three systems, the final volume 
containing the xenon (as estimated by measuring the advancement of ice surface) 
corresponds to concentrations of about 6--6.5\% which is consistent with the 
simulation temperatures of 5--7 K below ice freezing. However, no clathrate
hydrate is formed on the growing prismatic ice surface; instead, clathrate hydrate
forms via homogeneous nucleation when the concentration of xenon in
the bulk liquid reaches a threshold value.

\section{Discussion}

First, we would like to concisely recapitulate our key findings.

(\emph{i}) Hydration of xenon proceeds via the formation of clusters:
a xenon molecule plus its hydration shell (21.5 water molecules on
average). The cluster experiences diffusion as a whole, and the xenon
molecule in its center librates with a characteristic time of $\sim$2.2 ps and
an amplitude (squared) of \ensuremath{\le} (0.122 nm)\textsuperscript{2}, 
with occasional hopping from cage to cage.
In the crystalline clathrate hydrate phase, the period and amplitude
are slightly smaller (2.0 ps, (0.1 nm)\textsuperscript{2}).

(\emph{ii}) With increasing xenon concentration, the diffusion rate
slightly decreases. Possible explanations include (1) collisions of
clusters and (2) aggregation of clusters in potential clathrate hydrate
nuclei (similar to methane ``blobs'' observed in a recent study \cite{2010jacobsonamorphous}).

(\emph{iii}) Homogeneous nucleation of crystalline clathrate hydrate
is observed already at concentrations several times smaller than in
bulk hydrate. The structure of the phase seen in simulations agrees
well with the sI structure typical for real xenon hydrate, and the
lattice parameter is in good agreement with the experimental figure
of 1.182 nm.

(\emph{iv}) Formation of crystalline clathrate hydrate in all cases
is observed to happen radically faster than homogeneous nucleation of ice crystals. The excess (compared
to the equilibrium concentration of dissolved gas at the temperature
of simulation) is released via clathrate growth and not via gas bubble
formation.

(\emph{v}) Upon clathrate formation, the volume of the systems experiences
a sharp drop.

(\emph{vi}) At the simulation pressure of 1 atm, heating the clathrate-containing
systems shows no strong dependence of clathrate hydrate decomposition
temperature (between 275 and 280 K) on the concentration of xenon
in the surrounding solution.

(\emph{vii}) Although xenon does accumulate in the low-density interfacial
quasi-liquid layer on the surface of ice, clathrate hydrate formation
is not observed at ice surface in the equilibrium. However, xenon
does affect the morphology of the surface, making it more rough, and
decreasing the melting temperature of ice according to the cryoscopic
law.

(\emph{viii}) Under nonequilibrium conditions, the presence of xenon
in the interfacial layer does not prevent ice growth. Clathrate hydrate
is formed when the concentration of xenon in the not-frozen-yet part
of the system builds up sufficiently for homogeneous nucleation to
occur, and this happens in the thickness of the liquid layer.

(\emph{ix}) From (1) the independence of clathrate hydrate melting
temperature (unlike nucleation temperature) of the concentration of
xenon, (2) the experimentally observed dependence of decomposition
pressure on the temperature,\cite{1974ewingdissociation} and (3)
localization of clathrate hydrate nucleation not in the low-density/high-concentration
surface layer but in the bulk of the liquid, we conclude that the
main thermodynamical driving force behind clathrate hydrate stability
is the negative volume change upon addition of solvated molecules
to the clathrate crystal (via the associated negative enthalpy change).

(\emph{x}) At temperatures below $\sim$255--250 K, the viscosity
of xenon solution becomes so high that it has been impossible to reach
full convergence in the dynamic properties of the systems during our
long (100 ns) simulation runs, implying possible vitrification.

(\emph{xi}) The interaction of xenon molecules in solution can be
viewed as the balance between hydrophobic attraction of xenon molecules
and entropy-driven ``hydrational'' interaction between xenon and
water. The latter pulls apart contact (and even solvent-separated,
including those forming the clathrate lattice) pairs of xenon molecules.
With decreasing temperature, its relative role increases, resulting
in an exotic amorphous metastable structure with a short-range order
distance of at least $\sim$2 nm resembling an ``atomic gel''. 

Supersaturation of water with xenon can produce a phase that is metastable
with respect to both homogeneous nucleation of xenon clathrate hydrate
and the nucleation of ice. Although it is poorly understood how this
phase could be realized in practice, the high solubility of xenon
in lipid membranes \cite{2012yamamotodiffusive,2013bookerbiophysical}
could produce such an effect during cooling, when the surplus xenon
dissolved in the membranes gets ejected into the surrounding water
via an entropy-driven mechanism similar to hydrophobic dissociation.
Quantitative assessment of this effect requires separate confirmation
by simulations and experiment. If true, this may be one of the explanations
behind the protective effect of xenon observed experimentally on biological
samples. \cite{2008shelegcardiac}

At the same time, the present study provides evidence against the possibility of ``ice-blocking''
behavior of xenon, where clathrate hydrate could form on the surface
of ice, stopping its growth. This is because xenon hydrate formation
favors higher densities, while the density of quasi-liquid layer on
the surface of ice is lower than the density of bulk water. However,
this also suggests that other hydrophobic gases  might be more useful
in this respect (for example, recent molecular dynamics simulations
observed nucleation of \emph{methane} clathrate hydrate at ice--water
interfaces \cite{2013pirzadehmolecular}).

Finally, any cryoprotective effects specific to the interaction of
xenon with biological objects, especially those containing large hydrophobic
regions (proteins, lipid membranes), still remain to be investigated.
Given xenon's drastic effects on living organisms, including anaesthesia
and even drug addiction therapy, such interactions may turn out crucial
for the explanation of the cryoprotective effects of xenon. We hope
that the physicochemical findings of the present work may assist future
investigations of the underlying causes of the biological action
of xenon.

\section{Conclusions}

In the present work, we have carried out molecular dynamics simulations
of xenon in water under both homogeneous and heterogeneous (\emph{i.e.},
on ice surface) conditions. We observed the homogeneous nucleation
of xenon clathrate hydrate, as well as an exotic metastable amorphous
phase at high xenon concentrations and low temperatures. Despite the
fact that xenon adsorbs at the ice surface preferentially to being
dissolved in the liquid water, no clathrate hydrate nucleation is
seen at the interface. We trace this result to the lower density of
the interfacial quasi-liquid layer, while clathrate hydrate formation
is associated with an increase of density.

Together with the observed independence of xenon clathrate hydrate
decomposition temperature on the concentration, on one hand, and the
known temperature dependence of its dissociation pressure, on the
other, this leads to the conclusion that the stability of the hydrate
is governed not by the concentration of xenon in the liquid phase,
but by the interplay of thermal instability of the clathrate lattice,
the entropy-driven hydrophobic association/dissociation force, and
the negative contribution to the enthalpy resulting from the negative
change in volume associated with xenon molecules going from the liquid
to clathrate phase. Our results also hint at an optimal pressure-temperature
combination, above which the clathrate dissociates due to thermal
motion (translational entropy of xenon molecules, hydrophobic association
into contact solute pairs), and below which it is ``pulled apart''
by hydrophobic dissociation forces (entropy difference between ``bulk''
water molecules and those in xenon hydration shells).

These findings are interesting both for understanding the effects
of xenon on the freezing of water, as well as the broader aspects
of clathrate hydrate formation mechanisms. Our study lays the groundwork
for future investigations of the cryobiological effects of xenon.




%

\end{document}